\documentclass[11pt]{article}
\usepackage[utf8]{inputenc}
\usepackage[T1]{fontenc}
\usepackage{mathptmx}
\usepackage[margin=1.1in]{geometry}
\usepackage{microtype}
\usepackage{hanging}
\usepackage{setspace}
\usepackage[hidelinks]{hyperref}
\usepackage{xurl}
\usepackage{textcomp}

\title{Recommended Selves: Authenticity and Algorithmic Filtering\thanks{Preprint. Please cite the published version: Brown, Étienne. 2026. ``Recommended Selves: Authenticity and Algorithmic Filtering.'' \emph{Journal of the American Philosophical Association} 12 (1): 15--34. \url{https://doi.org/10.1017/apa.2025.10009}}}
\author{Étienne Brown\\ \small University of Ottawa}
\date{}

\begin{document}
\maketitle

\begin{abstract}
\noindent By allocating their attention to pieces of content, algorithmic filtering shapes the daily behavior of billions of users when they interact with a digital platform. Beyond conditioning what we do, can recommendation algorithms influence who we are? This article suggests that they do. Specifically, I contend that recommender systems affect users' capacity to be their authentic selves in both positive and negative ways. I start by offering an account of authenticity that builds on two central concepts: volitional alignment and self-understanding. I then explain how algorithmic filtering works and impacts authenticity. While recommender systems frustrate users' second-order desires by relying on uninformative behavioral signals, they also facilitate self-understanding by inciting users to question their identity. I end by discussing how controllable and explainable recommenders would best enable users to be authentic.
\end{abstract}

\onehalfspacing

Recommendations play a subtle but crucial role in our everyday lives. We read books and watch TV shows our friends have recommended to us. We stay in neighborhoods favored by the authors of popular travel guidebooks. We eat at restaurants with a rating of at least 4.0 out of 5.0 stars. Beyond leisure, recommendations also form the background of our most important decisions. We study at well-ranked institutions and choose careers based on trusted friends' or relatives' beliefs that we would flourish as lawyers, mechanics, teachers, or computer scientists. It seems appropriate -- even rational -- for us to make important life decisions based on recommendations. Why would we not? Arguably, we are more likely to be happy if we accept a job offer from a manager that our co-workers describe as a ``nice guy'' than if we join a team run by a ``petty workplace tyrant.'' Our confidence also seems warranted when we date someone our friends have vetted. When we must choose, having access to recommendations usually comes as a relief. Those who know us well can help us decide what suits us. Moreover, if our past decisions and actions are essential to who we are, the recommendations that influence them also contribute to our self-constitution as persons. We constitute ourselves through the exercise of our free agency, but also through the recommendations of others. We are, in this critical sense, recommended selves.

Before the rise of artificial intelligence, recommendations were made by humans. Yet, the rapid development of digital technologies has profoundly transformed the nature of recommendation. Today, many suggestions we follow are the product of recommender systems: algorithms that take ``a large set of items'' and determine ``which of those to display to a user.'' (Thorburn, Bengani, and Stray 2022). As Seaver (2019: 431) explains, ``Algorithmic recommendation has settled deep into the infrastructure of online cultural life, where it has become practically inescapable.'' To grasp the increasing role that automated recommendations play in our lives, it suffices to draw up a short list of popular apps on which billions of users rely daily: Facebook, X, Instagram, TikTok, YouTube, Apple News, Amazon, Spotify, Netflix, and Tinder. Recommender systems are ``among the most visible success stories of AI in practice'' (Jannach et al. 2022: 105:1).

Phenomenologically speaking, recommender systems are algorithmic engines that determine what users see \emph{first} and \emph{next} on a digital platform. As attention is limited, controlling what users see amounts to exercising a new social power (Bucher 2018; Aytac 2023; Lazar 2023). The next tweet we read, the next song we hear, perhaps even the next person we date: all of these are influenced by recommender systems that ``model our individual preferences and direct our attention to content we're likely to engage with'' (Schuster and Lazar 2024). For instance, recommendation algorithms on social media platforms influence which communities users join, what emotions they feel, what beliefs they form, and how they behave. Consequently, philosophers have held these platforms responsible for contributing to widely discussed social problems, including the spread of online hate (Guiora and Park 2017), misinformation (Harris 2024), radicalization (Alfano, Carter, and Cheong 2018), political polarization (Aikin and Talisse 2018), and social media addiction (Allcott, Gentzkow, and Song 2022).

In this paper, I aim to make a related but distinct contribution to the philosophy of algorithmic recommendation. Specifically, I suggest that recommender systems impact their users' capacity to be their authentic selves in both positive and negative ways. To make this claim plausible, I begin by offering a minimal account of authenticity, a philosophical concept that is ``deeply fraught and riddled with controversy'' (Rings 2017). My account is minimal insofar as I do not pretend to have identified all the conditions that must be met for a person to live a fully authentic life. Instead, I focus on two dimensions of authenticity -- volitional alignment and self-understanding -- currently affected by algorithmic filtering (section 1). I then offer a brief technical account of recommender systems (section 2). The remainder of the paper takes up the task of explaining how such systems affect authenticity (Sections 3 and 4). In a sense, my discussion adds to the recent philosophical critique of algorithmic recommendation. In section 3, I offer a version of this critique by arguing that current recommenders frustrate their users' second-order desires by relying on uninformative behavioral signals. That said, my reflection is also cautiously optimistic. In section 4, I suggest that algorithmic recommendation can facilitate self-understanding by inciting users to question their identity. As we will see, however, the emancipatory potential of recommender systems will only be realized if they become controllable and explainable. Opaque filtering risks misleading users into making wrongful assumptions about themselves. I end by discussing how controllable and explainable recommenders would best enable users to be authentic (section 5).

\section{A Minimal Account of Authenticity}

Many expressions belonging to the lexical field of authenticity are ambiguous. In everyday language, we speak of people who are ``true to themselves'' and actions that reflect ``who they really are'' without needing to explain what this means. Philosophers too frequently employ figurative expressions to describe authenticity. Bernard Williams, for instance, once said that his life writings are unified by the ``idea that some things are in some real sense really you, or express what you and others aren't'' (Jeffries 2002). This echoes an idea that Charles Taylor attributes to Herder: ``There is a certain way of being human that is my way. I am called upon to live my life in this way, and not in imitation of anyone else's.'' (Taylor 1992: 28-29) Philosophers also use allegorical phrases when they discuss inauthenticity. As Varga and Guignon point out (2023), Heidegger's analysis of \emph{Dasein} is guided by the idea that a person can ``lose herself'' by unreflectively acting as a member of ``the herd.'' More recently, Harry Frankfurt (1988) has described persons as beings who have the power to make some of their desires more truly their own and withdraw themselves from others. When we succeed or fail in following the desires we reflectively endorse, Frankfurt argues, ``the satisfactions at stake are those which accrue to a person of whom it may be said that his will is his own'' and ``the corresponding frustrations are those suffered by a person of whom it may be said that he is estranged from himself'' (22). When we succumb to weakness of will, ``our hearts are at best divided, and they may even not be in what we are doing at all.'' (163)

Can we disambiguate the metaphorical language of authenticity? In this section, I offer a philosophical account of this concept. According to my perspective, authenticity is a multi-dimensional concept. This means that Alice can be more authentic than Bob along some axis of authenticity while simultaneously being less authentic than him on another axis of authenticity. In simpler terms, Alice can be more authentic than Bob \emph{in some sense} but less authentic than him \emph{in some other sense}. Such an idea will become more concrete once I start identifying dimensions of authenticity. Note also that authenticity is not the only philosophical concept that can be reasonably envisioned as multi-dimensional. Happiness is another good example. Imagine, for instance, that Rosa is often in a bad mood but derives an essential sense of purpose and self-worth from her daily work with autistic children. By contrast, Jamal is a happy-go-lucky investment banker who occasionally feels some existential angst when thinking that he spends most of his life making the rich richer. This angst never lasts for long and can easily be suppressed with a few pickleball games. Here, it seems appropriate to say that Jamal is happier than Rosa in some sense but that, in another sense, he is not. Surely, being joyous daily is relevant to whether one is generally happy, but so is one's thought that one\textquotesingle s work is meaningful and contributes to the well-being of others.

Like there are many dimensions of happiness, I suggest there are many dimensions of authenticity. In what follows, I describe two of them. Although I hint at what other dimensions might be missing from my minimal account, my remarks will suffice to help us understand how recommender systems affect their users' capacity to be their authentic selves in later sections of this article.

Frankfurt has adequately conceptualized a first dimension of authenticity. Let us call it \emph{volitional alignment}. Michael Rings (2017: 479) summarizes it as follows:

\begin{quote}
An agent conducts her life authentically insofar as her second-order and first-order volitions align---i.e., she acts on decisions that reflect motivations or commitments she wholeheartedly endorses. These choices are expressive of who she is in the sense that they express her undivided will.
\end{quote}

Such an account of authenticity builds on Frankfurt's influential distinction between first-order and second-order desires. While first-order desires are unreflective, second-order desires are those that we reflectively endorse. We spontaneously want things (first-order desires), but we also sometimes want to want them (second-order desires). In other words, we often adopt a reflective and evaluative stance toward our spontaneous desires. When we fail to follow our second-order desires, then we are unable to realize our most profound aspirations. This is one important sense in which people fail to be true to themselves.\footnote{Volitional alignment closely relates to the theory of authenticity defended by Brink (2003: 251), who writes that authenticity requires acting on the ideals that the agent reflectively and sincerely accepts at the time of the action.' Recently, Brett Karlan (2024) has used Brink's work on authenticity to argue that algorithmically aided decision-making is often inauthentic.} Here, a paradigmatic example is that of an addict who attempts but fails to overcome his addiction. As a second example, imagine a frustrated and disengaged lawyer who chose her career merely to please her parents. While the lawyer fulfills her professional obligations, she deeply regrets engaging in this lifepath; what she truly wants to be is a painter. The lawyer wants to be a lawyer -- sufficiently so, at least, to be one every day -- but she does not reflectively endorse this desire. Deep down, she knows that good lives are lived by people who have the courage to resist parental directives and forge a path of their own. To live an authentic life, the lawyer would have to quit her job and give her best shot at painting.

Volitional alignment plays an important role in self-constitution. As Frankfurt (1988: 170) writes, ``The person, in making a decision by which he identifies with a desire, constitutes himself.'' When we successfully align our first- and second-order desires, we take steps toward becoming our authentic selves. For instance, we stop smoking, spend less time on our phones, are kinder to administrative staff, quit our unfulfilling jobs, and perhaps even become painters. Of course, when we systematically fail to follow our second-order desires, we also gradually forge our character through what we perceive as bad habits. If we accept volitional alignment as a dimension of authenticity, then we must accept that succumbing to bad habits -- even systematically so -- is a way of \emph{not} being ourselves. In other words, volitional alignment makes any philosophical account of authenticity aspirational: the person that ``we really are'' is constituted by our second-order desires even when such desires fail to direct our will.

The second dimension of authenticity I wish to discuss is epistemic instead of conative. Simply put, authenticity is enhanced by \emph{self-understanding}. To live authentic lives, we not only need to follow our second-order desires, but we must also understand what these desires are or construct the right ones by gathering self-knowledge. To identify the root of her malaise, the frustrated lawyer first needs to appreciate that she does not reflectively endorse her desire to please her parents but identifies instead with that to forge her own path. Within the history of philosophy, the ideal of self-understanding has given rise to the metaphor of ``turning inward.'' To apprehend our second-order desires, one thing we can do is engage in introspection. This ``inner-sense model of authenticity'' is often associated with Rousseau, who envisioned his \emph{Confessions} and \emph{Rêveries} as ways to disclose to his readers the authentic self that he had discovered through self-examination (Rings 2017: 487). Yet, as Varga and Guignon (2023) explain, this introspective view of authenticity has attracted virulent criticism. For instance, Adorno (1973: 70) contends that it rests on a ``liturgy of inwardness,'' according to which people are self-transparent subjects whose deepest desires are fixed and can be discovered by the mind's inner light. In Richard Rorty's view, Nietzsche also criticized the idea that being authentic amounts to ``coming to know a truth which was out there (or in there) all the time.'' (1989: 27). Contrary to Rousseau, philosophical critics of introspection tend to associate authenticity with \emph{self-creation} more than \emph{self-discovery}. Their point is that we constitute ourselves by defining (as opposed to identifying) our second-order desires throughout our lifetime.

I see no need to dispute the idea that authenticity involves self-creation. Obviously, we are not born with second-order desires. As psychologists have argued, some preferences are only constructed when elicited (Slovic 1995). We can imagine, for instance, that the unfulfilled lawyer's frustration with her career only arose when a colleague asked her if she always saw herself working as one. Only by reflecting upon this question did she come to identify and, eventually, reflectively endorse her desire to be a painter. In a sense, her desire to be a painter was ``constructed'' following her conversation. Still, the Nietzschean critique of self-understanding appears overblown. Surely, authenticity involves a fair amount of soul-searching. Even if we envision desires as constructed, it seems that constructing ones that will lead me to live a happy life requires me to understand important things about myself: what environments are best suited to my personality, with whom I like spending time, what kind of challenges I like giving to myself, etc. This is at least the case if we envision authenticity as a eudaimonic ideal and believe that a person's second-order desires should derive from a vision of ``flourishing that is specifically appropriate to the particular individual in question'' (Rings 2017: 476). It would be strange to attempt to create out of myself a person better suited to life pursuits that I naturally seek to avoid. As Rings observes, ``it would be wrong to banish all epistemic matters of self-knowledge entirely from the realm of authentic self-fulfillment.'' That we construct desires or have the power to self-create does not spare us from understanding who we are.

That said, we can do justice to critics of the inner-sense model of authenticity by pointing out that introspection is but one tool that people use to acquire self-knowledge. In what follows, I will emphasize that we come to understand who we are in large part through the advice and recommendations of others. For instance, parents often have a perception of their children that, when tactfully communicated, can help the latter make important life decisions. Think of a mother who asks her proud and brash son: ``I know that you want to be an oncologist, but how would you feel breaking bad news to patients several times per month?'' or ``How would you react to your scientific expertise being continuously challenged by skeptical patients?'' To accommodate the Nietzschean critique, we can emphasize that authenticity is \emph{relational} as much as personal. Specifically, the picture we need to reject is one in which a person's self-understanding can only be ``the product of his presence to himself'' (Williams 2022: 178). On the contrary, friends and relatives who know us well are a crucial source of self-knowledge, and listening to their advice can help us reflectively endorse desires that will lead us to flourish. To be our authentic selves, it helps to consider the recommendations of others. As we will see, these recommendations are also most useful when they are informative; that is when we understand \emph{why} things are recommended to us.

A full philosophical account of authenticity surely contains additional dimensions beyond volitional alignment and self-understanding. For instance, one could argue that children are authentic in a sense I have not discussed: they spontaneously express their thoughts and desires with little care for social conventions. Unlike many adults, they touchingly inhabit the world without worrying and wondering, ``What will they think if I wear, do, or say \emph{this}?'' One could also claim that authenticity has a temporal dimension. To live an authentic life, it seems that a person must \emph{commit} to certain desires and projects: a person who makes new decisions about who she wants to become every week would probably not strike us as authentic. Lastly, Charles Taylor (1992) has argued that authenticity can only be envisioned as a moral ideal if the authentic person's second-order desires relate to publicly shared values. This allows us to make sense of the idea that a life spent providing free education or healthcare to members of marginalized communities is more authentic than one devoted to counting blades of grass. While I find these claims plausible, they will not be at the forefront of my discussion in what follows.

\section{Recommender Systems: An Overview}

Recommender systems are attention-allocation engines that ``help users to find items of interest in situations of information overload.'' (Jannach et al. 2022: 105:1). Considering the technological ease with which users can now create digital content, there is more content to be viewed than viewers available to consume it. As Tim Wu (2018: 548) underlines, in the digital public sphere, ``it is no longer speech itself that is scarce, but the attention of listeners.'' The primary goal of recommendation algorithms is to solve this problem: instead of drowning in a sea of digital content they would rather not see, users should be presented with speech, media, products, and profiles that are of interest to them. In the information age, recommendation algorithms filter content on a wide array of heavily used digital platforms, including social media sites used by billions of users daily. Considering the sheer magnitude and reach of social media platforms, my focus in what follows will be on the recommender systems that structure social media feeds.

As Narayanan (2023) notes, one mistake to avoid when describing social media platforms is to conceive of them as solely powered by recommendation algorithms. Such platforms rely on a suite of algorithms only some of which recommend content to users. Other algorithms process content, for instance, by automatically analyzing text or tagging images. The propagation algorithms that deliver targeted advertising or friend recommendations are also distinct from those that populate a user's feed with content generated by other users. When I speak of ``recommendation algorithms,'' I primarily have in mind those that deliver (i) user-generated content, (ii) targeted advertisements, and (iii) group or friend recommendations.

Thorburn, Bengani, and Stray (2022) provide a useful four-stage description of how these algorithms work. The first stage is that of content moderation, in which pieces of content belonging to the total set of items that a user could view are automatically removed or flagged for violating a platform's policies (also called ``community guidelines''). The second stage is candidate generation. For each user, ``the full set of items available on the platform (potentially millions) is efficiently filtered to a set of~500 or so~that are plausibly of interest to the user.'' Candidate generation is followed by ranking, during which each remaining item in the set is given a score based on how engaging it is predicted to be for the user. The last stage is re-ranking and focuses on relationships between items. For instance, if the best-ranked posts for a user all come from the same content creator, the recommender might re-rank them to ensure more diversity (which is typically instrumental to better long-term engagement). In summary, the final ranking of items (or ``slate'') is the recommender's attempt to determine what moderated items are most likely to be engaging for a given user in the present context. According to Thorburn, Bengani, and Stray (2022), a simplified representation of the value model used to fulfill this task looks like this:

\begin{quote}
Value = 1 \textsuperscript{.} Pr(like) + 3 \textsuperscript{.} Pr(comment) +10 \textsuperscript{.} Pr(Share) + 4 \textsuperscript{.} (expected dwell time) -- 6 \textsuperscript{.} Pr(fake account) -- 7 \textsuperscript{.} Pr(low-quality news) -- 10 \textsuperscript{.} Pr(clickbait)
\end{quote}

Here, the ``linear combination of engagement predictions'' -- that is, the set of probabilities that relate to likes, comments, shares, and dwell time -- can be described as the \emph{core} ranking algorithm (Cunningham et al. 2024). In contrast, the \emph{full} ranking algorithm typically includes probabilities that are not predictions but instead estimate the likelihood that an item belongs to a pre-defined category of content (e.g., fake account, low-quality news, or clickbait). In other words, ``no real platform optimizes solely for engagement'' (Thorburn, Bengani, and Stray 2022). Quality control is also part of the equation.

To rank items, most social media recommender systems attempt to answer the following question\emph{:} ``How did users similar to this user engage with posts similar to this post?'' (Narayanan 2023). Moreover, to establish similarity between users, the core ranking algorithm typically considers both demographic attributes (e.g., age, gender, geographical location, language, etc.) and networks (i.e., who follows whom or is friends with whom). That said, \emph{behavioral history} remains the weightiest factor. In other words, what you view on social media largely depends on what items users categorized as similar to you have already engaged with.\footnote{As Narayanan explains, similarity between items is distinct from similarity between users. While the latter relies on behavior history, the former is usually tied to content metadata (e.g., two videos are deemed similar if they both have the term ``cooking'' in their title).} Like many other machine-learning algorithms, social media recommenders are thus actuarial in nature (Jorgensen 2022; Doyle 2024). The structure of the inferences they make is roughly the following: considering that users with property \emph{x} have engaged with content of type \emph{y}, this other user with property \emph{x} is likely to engage with content of type \emph{y}.\footnote{Here, property \emph{x} can itself be defined in terms of demographic attributes, network belonging, behavioral history, or a combination thereof. This will be important in later sections.} As Doyle (2024) notes, this means that algorithmic recommendation functions by including ``data about other people from relevant reference classes.''

However, the actuarial reasoning followed by social media recommenders does not perfectly correspond to that of human actuaries. One relevant difference is that human actuaries' reasoning relies on socially meaningful categories with which people frequently self-identify (e.g., 23-year-old male drivers tend to have \emph{x} accidents per year; therefore, the premium for all 23-year-old male drivers will be \emph{y}). By contrast, recommenders might ``automatically infer demographics like age, gender, and race as a side-effect of looking for patterns in the data'' (Narayanan, 2023). In 2018, this happened on Netflix when black users reported that recommendations seemed intentionally tailored to their race; the movies suggested to them often contained several black characters (Zarum 2018). In this case, however, racial targeting was a non-intentional emergent effect of algorithmic recommendation: as always, the goal was to engage users by relying on whatever category was necessary. If race works well, then so be it, but if other more abstract categories prove more effective at carrying out the predictive task at hand, they will be the ones used by the recommender. As a result, ``machine-generated categories may not correspond to any known social representation.'' (Milano, Taddeo, and Floridi 2020: 962). For instance, similarity in age or gender might be less salient to a recommender than the fact that a group of users recently bought a novelty sweater (962). Importantly, the categories that drive recommendations are not transparent to users. As we will see, this has the unfortunate effect of keeping users guessing why they are recommended the content they see.

\section{Modelled Preferences and Frustrated Desires}

Two distinct bodies of literature can help us see how recommender systems threaten volitional alignment. First, some scholars argue that digital platforms powered by recommendation algorithms are fundamentally manipulative (Susser, Roessler, and Nissenbaum 2019; Benn and Lazar 2022). According to this perspective, manipulation amounts to a form of hidden influence that exploits users' vulnerabilities to better steer their behavior towards the manipulators' ends. To illustrate this point, Susser, Roessler, and Nissenbaum (2018: 6) discuss a leaked internal Facebook strategy document in which one can read that ``By monitoring posts, pictures, interactions and internet activity in real-time, Facebook can work out when young people feel `stressed', `defeated', `overwhelmed', `anxious', `nervous', `stupid', `silly,' `useless,' and a `failure.'\,'' Using this information, the company could then direct targeted advertisement at them, hoping that their insecurity would translate into clicks. As the authors note, such ``manipulative influences thwart people's capacity to form decisions they can recognize and endorse as their own'' (4). The idea is that manipulation results in volitional misalignment as it covertly influences users' behavior in ways they cannot reflectively endorse.

A second body of literature relates to digital addiction. For instance, Allcott, Gentzkow, and Song (2022: 2427) contend that using social media platforms leads to self-control problems as ``people consume more {[}digital media{]} today than they would have chosen for themselves in advance.'' The authors draw this conclusion from an experiment in which they allowed subjects to activate screen time limits that could not be easily overridden on their smartphones. They then found ``clear evidence that people have self-control problems'' (2427). Indeed, ``78 percent of participants set binding limits and continued using them through the final weeks of the experiment.'' They did so even if they were not offered any incentive. Overall, ``the experiment reduced screen time by 22 minutes per day (16 percent) over 12 weeks.'' Here, the link with the philosophical account of authenticity I have proposed above is quite straightforward: Frankfurt's paradigmatic example of volitional misalignment is that of an unwilling addict whom he describes as a ``passive bystander to the forces that move him'' as he fails to control his addiction (1988: 22). Whether this person is addicted to nicotine or to the dopamine hit that comes from viewing one's post go viral does not matter. All addictions threaten authentic agency by impairing people's ability to act according to their second-order desires.

Discussions of online manipulation and digital addiction point to a deeper philosophical problem that deserves to be discussed: recommender systems foster volitional misalignment by design as they rely on behavioral signals that poorly correlate with their users' second-order desires. As we have seen, signals like clicks, likes, comments, and dwell time are the critical drivers of algorithmic recommendation on social media platforms (Stray et al. 2022). In other words, what users are recommended on such platforms essentially depends on what they have done while connected to the internet. As many have noted, however, how people behave is a poor indicator of their second-order desires. In the 1970s, Amartya Sen (1973; 1977) famously emphasized this point by arguing that a person's preferences cannot successfully be inferred from their actions. This is because people's mental lives are complex and involve unobservable desires. For instance, whether a smoker is peacefully enjoying a well-deserved reward or feeling ashamed to have bought yet another pack is not something that an observer can tell by seeing them light up a cigarette. To infer people's underlying preferences, we need more than the ``mere observation of actual choices'' (Sen 1977, 342); what we need is ``other sources of information, including introspection and discussion.''

Such a critique easily applies to recommender systems. To see this, imagine that I scroll down a social media feed for a few seconds\{Citation\}, skipping over video \emph{x} and then stopping to watch video \emph{y} for twelve seconds. Both videos were designed to catch my attention, but only \emph{y} succeeded. As a result, I am likely to be recommended more videos of kind \emph{y} in the future. Here are some possible descriptions of the situation:

\begin{itemize}
\item
  I watched \emph{y} as I generally enjoy watching videos like \emph{y}, and I would like to be recommended similar videos in the future.
\item
  I only watched \emph{y} as it was eye-catching, and I have no preference between videos like \emph{x} and \emph{y}.
\item
  I only watched \emph{y} as it was eye-catching. I believe watching videos like \emph{y} is a waste of time, and I would rather be recommended videos like \emph{x}.
\item
  I only watched \emph{y} as it was eye-catching. I believe watching videos like \emph{x} and \emph{y} is a waste of time, and I would rather be recommended videos of kind \emph{z}.
\item
  I did not watch \emph{y.} I stopped browsing as my spouse was talking to me.
\item
  I hate-watched \emph{y}, and I do not want to be recommended videos like \emph{y} it in the future.
\end{itemize}

A recommender system that merely considers behavioral signals cannot distinguish between these possibilities. Furthermore, it makes recommendations based on forms of behaviors that -- like watch or dwell time -- barely count as actions or decisions. When a flashy video catches my eye, and I spend a few seconds viewing it, I arguably do not \emph{decide} to watch it. Nevertheless, ``in an interpretative move inherited from behaviorism,'' recommenders take such behavioral signals ``as more truthful than users' explicit ratings'' (Seaver 2019: 430).

To sum up, current recommenders cannot track their users' second-order preferences as they primarily rely on behavioral data from which such preferences cannot be inferred. As a result, they foster volitional misalignment by design. Schuster and Lazar (2024) forcefully illustrate this objection:

\begin{quote}
You might be susceptible to clickbait with titles like ``she was famous in the 90s, see what she looks like now'' or ``presidential announcement sparks `total outrage'.'' But you might well have a higher-order preference not to be so inclined, if you care about being productive, focusing on accurate and important information, and limiting screen time more than the instant gratification that too often drives your behavior.
\end{quote}

A recommender that optimizes for volitional alignment would need better sources of information than behavioral signals. In section 5, I will discuss several technological changes that could prevent recommenders from fostering volitional misalignment by modeling user preferences in a more sophisticated manner. Specifically, I will contend that an authenticity-enhancing recommender could hardly avoid explicitly asking its users what their recommendation preferences are. As I would now like to suggest, however, it would also explain to them \emph{why} they are exposed to the content that they see.

\section{Listening to Recommenders}

Let me open this section with a hypothetical scenario:

\begin{quote}
Hao is a bisexual man in his mid-forties. Since his early twenties, he has been in a monogamous relationship with his wife, Jing. As a result of being in a monogamous heterosexual relationship, Hao's sexual attraction to men has never played a vital role in his life. Some of his friends (including Jing) know about it, but others do not. His bisexuality has somehow faded into the background of his daily life. Like many, Hao is moderately active on social media. While spending time online, he realizes that a significant proportion of content that is recommended to him is targeted at members of the LGBTQ+ community. This makes his bisexuality more salient to him. ``Not only am I sexually attracted to men, he thinks, but I behave like a member of the LGBTQ+ community. That is why I see all this content.'' This experience arises in Hao a desire to present himself as a member of the LGBTQ+ community and to explore his bisexuality.
\end{quote}

As it turns out, such a scenario is less hypothetical than appears at first glance. Indeed, many social media users report feeling ``seen'' by recommendation algorithms. Consider, for instance, the story of BBC journalist Ellie House, who realized that she was bisexual after ``getting more and more recommendations for series with lesbian storylines, or bi characters'' on Netflix and being suggested a ``playlist described as `sapphic' on Spotify.'' (House 2023).

Through algorithmic filtering, Hao and House came to understand something important about themselves. Reflecting upon the content displayed to them, they envisioned the recommender's ``testimony'' as a valuable source of self-knowledge and used it to construct their social identity. This represents a significant way in which algorithmic filtering can help users become their authentic selves. Remember indeed that the account of authenticity I proposed above is relational, that is, undergirded by the claim that testimony from others is just as valuable as introspection when it comes to self-discovery. My suggestion is that this point also applies to algorithmic recommendations. Certainly, coming to see that I behave similarly to people belonging to category \emph{x} -- say, members of the LGTBQ+ community or people on the ASD spectrum -- is a significant way in which a person can come to understand who they are. When recommenders help us do this, they foster self-understanding in a valuable manner. Of course, there are important differences between human recommendations and automated ones. In the scenarios above, the recommender did not provide Hao and House with advice but simply drew their attention to specific pieces of content. It is only by reflecting upon their feed that Hao and House drew conclusions about themselves and decided to explore their bisexuality. Ultimately, the recommendation to behave in this manner came from themselves. Still, it was prompted by information that the recommender delivered to them. Moreover, what our data exhausts reveal will likely differ from what we consciously disclose to our friends and relatives. If this is so, then the information that algorithmic filtering provides can complement the recommendations we receive from our close ones. Having a hard talk with your sister or best friend is certainly an excellent way of holding up a mirror and gathering information about the ``real you,'' but the picture of you that they hold in their mind is likely incomplete. Different people -- or systems -- see different things in us.

Perhaps I am moving too quickly. Hao and House treated algorithmic filtering as a source of self-knowledge, but is this appropriate? Can recommenders not also pose risks to self-understanding? Recently, Casey Doyle (2024) has argued that there is something wrong with ``listening to algorithms.'' In his view, ``To outsource inquiry about our minds to algorithms is to surrender the job of making up our own minds, which underlies the practice of first-person authority.'' To illustrate this claim, Doyle draws our attention to the Amazon Halo health band, which makes predictions about its users' emotional states based on the tone of their voice. Imagine, for instance, that my Halo band informs me that I am angry. Doyle contends that it would be inappropriate for me to defer to it without any reflection (``\,`I'm angry,' I tell you, `my band has informed me so.'\,'')

Here is the reasoning that undergirds Doyle's argument. As we have seen, machine-learning algorithms make actuarial generalizations about groups to which users belong. However, ``to rely on actuarial reasoning about one's own or another's mind fails to respect the fact that it is up to the subject herself what she believes, desires, and intends to do'' (Doyle 2024). To make his case, Doyle draws from the work of scholars who worry that using predictive algorithms in criminal law amounts to failing to treat people as free autonomous agents.\footnote{See Jorgensen (2022) for a discussion.} Again, this is because such algorithms rely on actuarial reasoning: ``The worry is that decisions based on this sort of evidence amount to `guilt by membership in a reference class.'\,'' According to Doyle, a similar worry applies in the Halo band case. As he puts it, ``In both cases, we feel entitled to be treated as an exception to generalizations about groups of which we are members.''

Yet, there are relevant dissimilarities between the three cases at hand: (i) Hao and House's reliance on social media feeds to gather self-knowledge, (ii) the Amazon Halo band, and (iii) the use of algorithmic predictions in criminal law. First, neither Hao nor House \emph{defers} to a recommender system without reflection. They do not believe that they are bisexual because the algorithm has told them so. Instead, they use algorithmic recommendations to \emph{raise questions} that the recommender cannot answer for them: ``Could I be bisexual?''; ``Am I more similar to other members of the LGBTQ+ community than I thought?''; ``Should I explore my bisexuality?'' etc.

This difference matters. As Doyle (2024) notes, his argument only applies to ``cases of pure deference'' in which a person takes an algorithmic recommendation or decisions ``at face value, by contrast with taking it into consideration along with the rest of one's evidence.'' Second, contrary to Doyle, contemporary decision theorists have argued that engaging in actuarial reasoning to gather information about oneself and make important life decisions is a rational way to behave (Pettigrew 2019: 136-138). For instance, to decide whether I should have a child, it seems appropriate to gather information about whether people whose lives and personalities share many similarities with my own have a positive experience of parenthood.\footnote{This claim is debated in the philosophical literature on transformative experience. See the exchange between Paul (2016) and Pettigrew (2019).} If this is so, then similar cases of actuarial reasoning are arguably not inappropriate, and the fact that recommenders rely on such reasoning is not a sufficient reason to believe that they cannot facilitate self-discovery. Hao and House can go ahead and think that, based on the behavioral history they share with members of the LGBTQ+ community, they might enjoy exploring their sexuality.

There is, however, a second reason to worry that recommenders might prevent Hao and House from gathering knowledge about themselves. Their reconstruction of why they are recommended LGBTQ+ content might be a figment of their imagination. Consider Hao's case more closely. Certainly, the recommender's ranking of items on his social media feed \emph{might} be driven by the behavioral history he shares with other members of the LGBTQ+ community, but there is no way for him to tell. As we have seen in the case of Black movie recommendations on Netflix, racial or LGBTQ+ targeting is, at best, a non-intentional emergent effect of algorithmic recommendation optimized for engagement. In fact, the recommender system might be recommending LGBTQ+ content to him for reasons that have little to do with his own presumed queer behavior. Perhaps, for instance, the recommender ``knows'' that Hao and his wife live in the Castro -- a San Francisco neighborhood densely populated by LGBTQ+ people -- and displays LGBTQ+ content because of their geographical location. As Schuster and Lazar (2024) note, ``recommender systems use deep learning with artificial neural networks to discover highly complex patterns among users' behavior, features of content, and many other factors, which elude human analysis.'' As a result, there is no way for Hao to understand why he is being recommended LGBTQ+ content, and his belief that he behaves like a member of the LGBTQ+ community might be false.

Will algorithmic recommendations necessarily prevent users from understanding who they are because it is fundamentally opaque? Perhaps not. After all, Hao already knows that he is bisexual and merely uses algorithmic recommendations to reflect upon what it means for him to be bisexual. He can also answer this question even if it turns out that he was primarily recommended LGBTQ+ content because of his geographical location. Yet, this should not lead us to understate how valuable it is for people to have access to \emph{informative} recommendations when they attempt to make decisions that will lead them to flourish. We not only want to receive recommendations but also understand what motivates them. Imagine, for instance, that my partner asks me whether she should accept a new job. ``You should,'' I respond without elaborating. My recommendation is valuable to some extent. I have known my partner for fourteen years, and we have reasons to take each other's advice seriously. Still, it would be better for me to respond, ``You should; this is a managerial position, and I have always thought you have the qualities needed to be a great manager,'' or ``You should; you'll cut your commute time in half.'' When I make recommendations \emph{and} explain why, my partner is in a better position to evaluate such recommendations and make decisions accordingly. That I see her flourishing in a managerial position might help her form the second-order desire to become one. By comparison, she might see my comment about her commute time as insignificant. In general, we are better positioned to gather self-knowledge when we understand why certain things are recommended to us, and this knowledge is useful when we must make important decisions.

Again, my suggestion is that this point applies to both human and algorithmic recommendations. To see this, consider the following variation on the Hao case:

\begin{quote}
Farah is a Muslim woman who was raised in a conservative religious family. She has always felt some discomfort with her relatives' conservative beliefs. While living abroad, Farah has befriended many progressive Muslims and now sees herself as one. Like many, Farah is moderately active on social media. While spending time online, she realizes that a significant proportion of content that is recommended to her is targeted at conservative Muslim women. This annoys her and makes her identity as a progressive more salient to her. ``I thought I was a progressive Muslim, she thinks, but I behave like a conservative one.'' This experience leads her to redouble her efforts to surround herself with progressive friends and limit contact with her conservative relatives.
\end{quote}

As we did for Hao, we can easily picture a situation where Farah's explanation of why she is recommended conservative content is a figment of her imagination. Let us say, for instance, that she lives in Whitechapel -- a neighborhood in London heavily populated with Muslims -- and geographical location is the primary reason why she is recommended such content. Like Hao, Farah might still flourish by redoubling her efforts to surround herself with progressive Muslims. Yet, such efforts might also make her unhappy. Regardless of their outcome, they are also motivated by the wrong reasons: there is no point in attempting to overcome a part of yourself you dislike if that part is not there in the first place. Suppose Farah understood that the conservative content she saw was recommended to her based on her location rather than her behavior. In that case, she might not have prioritized her family relationships over her friendships. While Hao's case illustrates the emancipatory potential of algorithmic filtering, Farah's example helps us see that only \emph{explainable} recommendations can fully realize this potential. Algorithmic filtering naturally makes users wonder why they are recommended what they see, but obscure recommendations risk misleading them into entertaining delusions about themselves. This claim is supported by the scientific literature on algorithmic folk theories, which emphasizes that users spontaneously develop their own, often inaccurate, theories of how opaque recommenders work to make sense of their social media experiences (Eslami et al. 2016; DeVito et al. 2018; Karizat et al. 2021). In summary, algorithmic filtering can facilitate self-understanding; however, opacity remains a significant obstacle.

\section{Optimizing for Authenticity}

The practical upshot of my reflection is that we need better recommenders to be our authentic selves. First, recommender systems could facilitate volitional alignment by offering users more control over what they see. Proposals in this direction range from minor adjustments to the attention economy to more ambitious reforms enabled by large language models. Regarding minor changes, digital platforms could solicit more feedback from users by asking them to complete surveys (``Would you like to see more diverse content?'' ``Are you using social media to keep up with the news?'' etc.) (Cunningham et al. 2024). They could then use survey responses to personalize filtering. Of course, the effectiveness of such measures will depend on several factors, including whether users are being asked the right questions, how often they respond to surveys, and to what degree engagement optimization is modified to reflect users' explicit preferences. Alternatively, users could have a better opportunity to choose between different algorithmic feeds. For instance, Fukuyama (2021) has recently defended the implementation of ``middleware,'' that is, content-curation services that would offer users ``feeds curated according to alternate ranking, labeling, or content-moderation rules'' (Keller 2021: 168). The goal is to create a new market of digital services that users could ``plug in'' their social media accounts to personalize their algorithmic feed: ``In place of a nontransparent algorithm built into the platform, you could decide to use a filter provided by a nonprofit coalition of universities that would vouch for the reliability of data sources'' (Fukuyama 2021: 42). A more radical proposal would be to replace current recommendation algorithms with conversational recommenders, that is, software systems that support their ``users in achieving recommendation-related goals through a multi-turn dialogue'' (Jannach et al. 2022: 105:3). In broad strokes, the idea is to integrate algorithmic recommendations with large language models, enabling users to continuously interact with their recommender (``Could you make it so that I see more LGBTQ+ content, please?''). Relatedly, Schuster and Lazar (2024) favor the creation of LLM-powered ``generative agents'' that would function ``as the executive control center of a multifaceted system'' and serve as ``digital content sommelier or attention guardian.'' Digital assistants could be built into the operating systems of personal computers, gather information about the kind of experience users seek, and eliminate the need to log in to social media by directly providing them with content that matches their explicit preferences. One central challenge recommender controls face is that users often do not wish to express their preferences. As Stray and his collaborators explain (2022), ``Even when controls are provided, many users {[}\ldots{]}~simply don't see the value in engaging with them.'' As a result, a passive experience of algorithmic recommendation ``remains the default.'' However, it is worth noting that this empirical finding only applies to controls currently implemented by social media platforms (e.g., ``See fewer posts like this''), not to more ambitious proposals, such as middleware, conversational recommenders, or generative agents.

Yet, as we have seen, volitional alignment is not all there is to authenticity. The challenge for Farah was not that algorithmic recommendation frustrated her second-order desires, but rather that it misled her into making a wrongful assumption about herself. A straightforward solution to this problem is the implementation of explainable recommenders, that is, recommendation engines that explain to users the motivations behind their recommendations (Zhang and Chen 2020). Such recommenders are based on ``explainability-aware ML techniques'' and can be categorized into two main groups (Marconi, Matamoros A., and Epifania 2022). First, some black box recommenders still provide users with post hoc explanations of their output. Yet, they do so without offering ``an in-depth understanding of the underlying algorithm.'' Second, so-called white box recommenders directly incorporate interpretable machine learning models into algorithmic filtering. Often called ``explainability-by-design,'' this second approach avoids the challenges of interpreting opaque recommenders already in use (Stray et al. 2022).

Unfortunately, explainable recommenders face many challenges. First, discussions of explainable AI tend to focus on predictions that lead to the most straightforward exercise of algorithmic power over individuals (GDPR 2016; Vredenburgh 2022). Think, for instance, of credit-scoring algorithms (Hurlin, Pérignon, and Saurin 2022) or risk predictions used to justify the denial of pre-trial release or parole (Angwin et al. 2016). As these predictions relate to the distribution of societal benefits and burdens, they pose a clear risk to fairness. By way of contrast, recommendation algorithms impact their users in more subtle ways. As a result, they are less frequently the object of philosophical reflections that focus on transparency.\footnote{For instance, neither Vredenburgh (2022) nor Karlan (2024) discuss cases that involve recommendation algorithms used on social media (although the latter does discuss a case that involves the use of the Google search algorithm).} Second, the technical means necessary to explain the most sophisticated recommender systems are still missing: ``With the rise of increasingly complex machine learning models, it has also become increasingly difficult to give intuitive and understandable explanations of why a user received a specific recommendation'' (Stray et al. 2022).

A third challenge to developing explainable recommenders is that there is no explicit agreement on (i) the conditions that must be met for an AI model to count as explainable and (ii) why explainability is valuable (especially when no apparent risk to fairness is at play). The preceding discussion offers elements of an answer to these question. Understanding why people or algorithms recommend things to us can help us form a more accurate picture of who we are. Narayanan's (2023) characterization of social media recommenders also provides a partial answer to the question of what explainable recommendations should consist of. To reach a higher level of self-understanding, users should be explained what is meant by ``similar'' in expressions like ``similar users'' and ``similar content.'' For Hao, an explanation such as ``You're viewing LGBTQ+ content as \emph{users like you} have engaged with similar content'' is valueless. To calibrate his reaction to the recommender's testimony, he must know whether the expression ``users like you'' is understood in behavioral or geographical terms (or, if both factors are relevant, how weighty each of them is). In short, explainable recommenders must allow the average user to comprehend which similarity relationships the model constructs. This is at least the case if the value of explanations partly derives from their ability to foster self-understanding.

Lastly, there is a potential tension between the quest for self-understanding and the desire for users to exercise greater control over recommender systems. In section 1, I suggested that authenticity not only requires alignment between our current first- and second-order desires, but also the discovery of what our second-order desires are. Is there not a risk that offering users more control~over recommenders will hinder self-understanding by preventing them from identifying their second-order desires? Imagine a world in which Ellie House exercises so much control over her algorithmic feeds that she is never recommended the ``sapphic'' playlists that sparked a reflection on her sexual orientation. Perhaps, for instance, redesigned recommenders never suggest a playlist before explicitly asking her what she would like to listen to, in which case she is unlikely to request sapphic music. In general, will implementing controls not diminish the likelihood of encounters that prompt users to question or revise their current desires?\footnote{I thank an anonymous reviewer for helping me articulate this tension.}

The risk is real, but it can be mitigated. In House's case, what sparked self-understanding is the realization that a recommender system had categorized her as \emph{behaving like} members of a particular group. In general, user controls only risk hindering self-understanding if they render categorizations that involve the construction of similarity relationships obsolete or invisible. For instance, if I tell a conversational recommender that I want to be shown ten random cooking articles from The New York Times daily, the recommender does not need to construct similarity relationships to accomplish this task. That said, we can design user controls that (i) do not eliminate the construction of similarity relationships and (ii) make them explicit. For instance, survey questions could be phrased to help users see what similarity relationships recommenders construct (``We have noticed that you often engage with content popular with the LGBTQ+ community. Would you like to see more content of this kind?''). Furthermore, conversational recommenders could be programmed to construct such relationships and provide users with information about them (``Would you like to know more about the characteristics and interests of users who, like you, enjoy The New York Times cooking articles?''). In other words, at least some user controls can provide users with opportunities for self-discovery. Yet, the potential tension between volitional alignment and self-understanding should not be dismissed too quickly. If we forget that recognizing \emph{who we behave like} can facilitate self-understanding, there is a risk that we will design controls that prevent users from reaching this kind of realization.

\section{Conclusion}

Why look at recommender systems through the philosophical prism of authenticity? Should we not avoid the ``jargon of authenticity'' (Adorno 1973), with all its metaphors and ambiguity? Is there no philosophical concept that better allows us to understand how algorithmic filtering affects us? Like many other concepts, authenticity has a rich philosophical history despite being considerably difficult to define. As the existence of dozens of self-help books on the subject suggests, it is also a notion to which people easily relate in everyday reflection (Thacker 2016; Kentgen 2018; Nadrich 2019; Stahl 2020). Simply put, many people care about being authentic. At a time when discussions of artificial intelligence are omnipresent within the philosophical discipline, I find it valuable to relate the ``jargon of AI'' to concepts that play an essential role in everyday philosophical thinking. This is but one way philosophers can democratize the prohibitively technical discussions of AI in which they sometimes engage (for good reasons). This is a tall order, but I hope it helps the reader understand what I have attempted to do in this paper. If my suggestions are plausible, a central reason why we should be dissatisfied with current recommenders is that they meaningfully prevent us from being ourselves. Yet, they also have the potential to help us understand who we are.

\section*{References}
\begin{hangparas}{2em}{1}
\setlength{\parskip}{6pt}

Adorno, Theodor W. 1973. \emph{The Jargon of Authenticity}. Evanston, Ill: Northwestern University Press.

Aikin, Scott F., and Robert B. Talisse. 2018. \emph{Why We Argue (and How We Should): A Guide to Political Disagreement in the Age of Unreason}. Second edition. New York, NY: Routledge.

Alfano, Mark, J. Adam Carter, and Marc Cheong. 2018. ``Technological Seduction and Self-Radicalization.'' \emph{Journal of the American Philosophical Association} 4 (3): 298--322. \url{https://doi.org/10.1017/apa.2018.27}.

Allcott, Hunt, Matthew Gentzkow, and Lena Song. 2022. ``Digital Addiction.'' \emph{American Economic Review} 112 (7): 2424--63. \url{https://doi.org/10.1257/aer.20210867}.

Angwin, Julia, Jeff Larson, Surya Mattu, and Lauren Kirchner. 2016. ``Machine Bias.'' ProPublica. 2016. \url{https://www.propublica.org/article/machine-bias-risk-assessments-in-criminal-sentencing}.

Aytac, Ugur. 2023. ``Global Political Legitimacy and the Structural Power of Capital.'' \emph{Journal of Social Philosophy} 54 (4): 490--509. \url{https://doi.org/10.1111/josp.12464}.

Benn, Claire, and Seth Lazar. 2022. ``What's Wrong with Automated Influence.'' \emph{Canadian Journal of Philosophy} 52 (1): 125--48. \url{https://doi.org/10.1017/can.2021.23}.

Brink, David O. 2003. ``Prudence and Authenticity: Intrapersonal Conflicts of Value.'' \emph{The Philosophical Review} 112 (2): 215--45. \url{https://doi.org/10.1215/00318108-112-2-215}.

Brown, Étienne. 2018. ``Propaganda, Misinformation, and the Epistemic Value of Democracy.'' \emph{Critical Review} 30 (3--4): 194--218. \url{https://doi.org/10.1080/08913811.2018.1575007}.

Bucher, Taina. 2018. \emph{If...Then: Algorithmic Power and Politics}. New York: Oxford University Press.

Cunningham, Tom, Sana Pandey, Leif Sigerson, Jonathan Stray, Jeff Allen, Bonnie Barrilleaux, Ravi Iyer, Smitha Milli, Mohit Kothari, and Behnam Rezaei. 2024. ``What We Know About Using Non-Engagement Signals in Content Ranking.'' \url{https://doi.org/10.48550/ARXIV.2402.06831}.

DeVito, Michael A., Jeremy Birnholtz, Jeffery T. Hancock, Megan French, and Sunny Liu. 2018. ``How People Form Folk Theories of Social Media Feeds and What It Means for How We Study Self-Presentation.'' In \emph{Proceedings of the 2018 CHI Conference on Human Factors in Computing Systems}, 1--12. Montreal QC Canada: ACM. \url{https://doi.org/10.1145/3173574.3173694}.

Doyle, Casey. 2024. ``Listening to Algorithms: The Case of Self‐knowledge.'' \emph{European Journal of Philosophy}, January, ejop.12923. \url{https://doi.org/10.1111/ejop.12923}.

Eslami, Motahhare, Karrie Karahalios, Christian Sandvig, Kristen Vaccaro, Aimee Rickman, Kevin Hamilton, and Alex Kirlik. 2016. ``First I `like' It, Then I Hide It: Folk Theories of Social Feeds.'' In \emph{Proceedings of the 2016 CHI Conference on Human Factors in Computing Systems}, 2371--82. San Jose California USA: ACM. \url{https://doi.org/10.1145/2858036.2858494}.

Frankfurt, Harry G. 1988. \emph{The Importance of What We Care about: Philosophical Essays}. Cambridge {[}England{]}\,; New York: Cambridge University Press.

Fukuyama, Francis. 2021. ``Making the Internet Safe for Democracy.'' \emph{Journal of Democracy} 32 (2): 37--44. \url{https://doi.org/10.1353/jod.2021.0017}.

Guiora, Amos, and Elizabeth A. Park. 2017. ``Hate Speech on Social Media.'' \emph{Philosophia} 45 (3): 957--71. \url{https://doi.org/10.1007/s11406-017-9858-4}.

Harris, Keith Raymond. 2024. \emph{Misinformation, Content Moderation, and Epistemology: Protecting Knowledge}. 1st ed. New York: Routledge. \url{https://doi.org/10.4324/9781032636900}.

House, Ellie. 2023. ``Netflix: How Did It Know I Was Bi before I Did?,'' August 12, 2023. \url{https://www.bbc.com/news/technology-66472938}.

Hurlin, Christophe, Christophe Pérignon, and Sébastien Saurin. 2022. ``The Fairness of Credit Scoring Models.'' \url{https://doi.org/10.48550/ARXIV.2205.10200}.

Jannach, Dietmar, Ahtsham Manzoor, Wanling Cai, and Li Chen. 2022. ``A Survey on Conversational Recommender Systems.'' \emph{ACM Computing Surveys} 54 (5): 1--36. \url{https://doi.org/10.1145/3453154}.

Jorgensen, Renée. 2022. ``Algorithms and the Individual in Criminal Law.'' \emph{Canadian Journal of Philosophy} 52 (1): 61--77. \url{https://doi.org/10.1017/can.2021.28}.

Karizat, Nadia, Dan Delmonaco, Motahhare Eslami, and Nazanin Andalibi. 2021. ``Algorithmic Folk Theories and Identity: How TikTok Users Co-Produce Knowledge of Identity and Engage in Algorithmic Resistance.'' \emph{Proceedings of the ACM on Human-Computer Interaction} 5 (CSCW2): 1--44. \url{https://doi.org/10.1145/3476046}.

Karlan, Brett. 2024. ``Authenticity in Algorithm-Aided Decision-Making.'' \emph{Synthese} 204 (3): 93. \url{https://doi.org/10.1007/s11229-024-04716-7}.

Keller, Daphne. 2021. ``The Future of Platform Power: Making Middleware Work.'' \emph{Journal of Democracy} 32 (3): 168--72. \url{https://doi.org/10.1353/jod.2021.0043}.

Kentgen, Lisa. 2018. \emph{An Authentic Life: Five Foundations of Authenticity \& Purpose}. 1st edition. New York, NY: Stryder Press.

Lazar, Seth. 2023. ``Communicative Justice and the Distribution of Attention.'' \emph{Knight First Amendment Institute at Columbia University Essays and Scholarship}, October 10, 2023. \url{https://knightcolumbia.org/content/communicative-justice-and-the-distribution-of-attention}.

Marconi, Luca, Ricardo A. Matamoros A., and Francesco Epifania. 2022. ``Discovering the Unknown Suggestion: A Short Review on Explainability for Recommender Systems.'' \emph{Proceedings of the 2nd Italian Workshop on Artificial Intelligence and Applications for Business and Industries}.

Milano, Silvia, Mariarosaria Taddeo, and Luciano Floridi. 2020. ``Recommender Systems and Their Ethical Challenges.'' \emph{AI \& SOCIETY} 35 (4): 957--67. \url{https://doi.org/10.1007/s00146-020-00950-y}.

Nadrich, Ora. 2019. \emph{Live True: A Mindfulness Guide for Authenticity}. Los Angeles, CA: Institute For Transformational Thinking Press.

Narayanan, Arvind. 2023. ``Understanding Social Media Recommendation Algorithms.'' Knight First Amendment Institute. March 9, 2023. \url{http://knightcolumbia.org/content/understanding-social-media-recommendation-algorithms}.

Paul, L. A. 2016. \emph{Transformative Experience}. Oxford: Oxford University Press.

Pettigrew, Richard. 2019. \emph{Choosing for Changing Selves}. First edition. Oxford\,; New York, NY: Oxford University Press.

``Regulation (EU) 2016/679 of the European Parliament and of the Council of 27 April 2016 on the Protection of Natural Persons with Regard to the Processing of Personal Data and on the Free Movement of Such Data, and Repealing Directive 95/46/EC (General Data Protection Regulation).'' 2016. 2016. \url{http://data.europa.eu/eli/reg/2016/679/oj}.

Rings, Michael. 2017. ``Authenticity, Self-Fulfillment, and Self-Acknowledgment.'' \emph{The Journal of Value Inquiry} 51 (3): 475--89. \url{https://doi.org/10.1007/s10790-017-9589-6}.

Rorty, Richard. 1989. \emph{Contingency, Irony, and Solidarity}. Cambridge\,; New York: Cambridge University Press.

Schuster, Nick, and Seth Lazar. 2024. ``Attention, Moral Skill, and Algorithmic Recommendation.'' \emph{Philosophical Studies}, January. \url{https://doi.org/10.1007/s11098-023-02083-6}.

Seaver, Nick. 2019. ``Captivating Algorithms: Recommender Systems as Traps.'' \emph{Journal of Material Culture} 24 (4): 421--36. \url{https://doi.org/10.1177/1359183518820366}.

Sen, Amartya. 1973. ``Behaviour and the Concept of Preference.'' \emph{Economica} 40 (159): 241--59. \url{https://doi.org/10.2307/2552796}.

Sen, Amartya K. 1977. ``Rational Fools: A Critique of the Behavioral Foundations of Economic Theory.'' \emph{Philosophy \& Public Affairs} 6 (4): 317--44.

Slovic, Paul. 1995. ``The Construction of Preference.'' \emph{American Psychologist} 50 (5): 364--71. \url{https://doi.org/10.1037/0003-066X.50.5.364}.

Stahl, Stefanie. 2020. \emph{The Child in You: The Breakthrough Method for Bringing out Your Authentic Self}. New York: Penguin Books.

Stray, Jonathan, Alon Halevy, Parisa Assar, Dylan Hadfield-Menell, Craig Boutilier, Amar Ashar, Lex Beattie, et al. 2022. ``Building Human Values into Recommender Systems: An Interdisciplinary Synthesis.'' \url{https://doi.org/10.48550/ARXIV.2207.10192}.

Jeffries, Stuart. 2002. ``The Quest for Truth. An Interview with Bernard Williams.'' \emph{The Guardian}, November 30, 2002, sec. Books. \url{https://www.theguardian.com/books/2002/nov/30/academicexperts.highereducation}.

Susser, Daniel, Beate Roessler, and Helen F. Nissenbaum. 2019. ``Online Manipulation: Hidden Influences in a Digital World.'' \emph{Georgetown Law Review} 4 (1). \url{https://doi.org/10.2139/ssrn.3306006}.

Taylor, Charles. 1992. \emph{The Ethics of Authenticity}. Cambridge, Mass: Harvard University Press.

Thacker, Karissa. 2016. \emph{The Art of Authenticity: Tools to Become an Authentic Leader and Your Best Self}. Hoboken, New Jersey: Wiley.

Thorburn, Luke, Priyanjana Bengani, and Jonathan Stray. 2022. ``How Platform Recommenders Work.'' \emph{Understanding Recommenders} (blog). November 23, 2022. \url{https://medium.com/understanding-recommenders/how-platform-recommenders-work-15e260d9a15a}.

Varga, Somogy, and Charles Guignon. 2023. ``Authenticity.'' In \emph{The Stanford Encyclopedia of Philosophy}, Summer 2023. \url{https://plato.stanford.edu/archives/sum2023/entries/authenticity/}.

Vredenburgh, Kate. 2022. ``The Right to Explanation*.'' \emph{Journal of Political Philosophy} 30 (2): 209--29. \url{https://doi.org/10.1111/jopp.12262}.

Wu, Tim. 2018. ``Is the First Amendment Obsolete?'' \emph{Michigan Law Review}, no. 117.3, 547. \url{https://doi.org/10.36644/mlr.117.3.first}.

Zarum, Lara. 2018. ``Some Viewers Think Netflix Is Targeting Them by Race. Here's What to Know.'' \emph{The New York Times}, October 23, 2018, sec. Arts. \url{https://www.nytimes.com/2018/10/23/arts/television/netflix-race-targeting-personalization.html}.

Zhang, Yongfeng, and Xu Chen. 2020. ``Explainable Recommendation: A Survey and New Perspectives.'' \emph{Foundations and Trends® in Information Retrieval} 14 (1): 1--101. \url{https://doi.org/10.1561/1500000066}.
\end{hangparas}

\end{document}